\documentclass[twocolumn,prl,10pt,aps,floatfix]{revtex4-1}
\usepackage{graphicx}
\usepackage{amsmath}
\usepackage[colorlinks,linkcolor=blue,citecolor=blue,urlcolor=blue]{hyperref}
\usepackage{color}

\begin{document}
\title{Reactive Hall and Edelstein effects in a tight-binding model with 
spin-orbit coupling}
\author{Emmanouil K. Kokkinis$^{1}$, Joseph J. Betouras$^2$, 
and Xenophon Zotos$^{3,4}$}
\affiliation{$^1$ 
School of Physics and Astronomy, University of Minnesota, 
Minneapolis, MN 55455, USA}
\affiliation{$^2$ Department of Physics, Loughborough University, 
Loughborough LE11 3TU, United Kingdom} 
\affiliation{$^3$Department of Physics,
University of Crete, 70013 Heraklion, Greece}
\affiliation{$^4$
Max-Planck-Institut f\"ur Physik Komplexer Systeme, 01187 Dresden, Germany}

\date{\today}

\begin{abstract}
The reactive Hall constant $R_H$, described by reactive 
(non-dissipative) conductivities, is analyzed within linear response theory
in the presence of spin-orbit interaction.
Within a two dimensional tight-binding model the effect of van Hove 
singularities is studied. 
Along the same line a formulation of the Edelstein constant is 
proposed and studied as a function of coupling parameters and 
fermion filling.
\end{abstract}
\pacs{PACS numbers: 71.27.+a, 71.10.Fd, 72.15.Gd}

\maketitle

\section{Introduction}
The spin-orbit interaction (SOI) \cite{rashba1,rashba2} 
plays a prominent role in the field of spintronics.
It has been extensively studied as a key mechanism in the anomalous Hall 
effect \cite{nagaosa}, the spin Hall effect \cite{dyakonov,hirsh,kato,sinova}, 
the induced magnetoelectric torque and edge currents \cite{balseiro, sigrist}, 
the Edelstein effect \cite{edelstein,mertig} 
to name just a very few. 

From a different perspective, in a seminal work \cite{kohn} W. Kohn brought 
attention to the reactive response of an electronic system,   
as criterion of the Mott metal-insulator transition, 
given by the imaginary part of the conductivity, 
$\sigma''(\omega\rightarrow 0)=2D/\omega$. 
The prefactor $D$, referred to as ''Drude" weight in numerous recent 
theoretical studies and also  equal to 
the weight of the zero-frequency $\delta$-function in the regular part 
of the conductivity, is finite in a non-interacting gapless system 
without disorder at all temperatures or a gapless 
strongly correlated integrable one \cite{znp}.  
In the presence of scattering (e.g. phonons, disorder) the $\delta$-function 
broadens to a peak of width $1/\tau$ 
(where $\tau$ is a characteristic scattering time)
and integrated weight $D$ which is experimentally studied.

Along the same line, a formulation of the reactive Hall  response
was proposed \cite{znlp} in order to address the problem of Hall constant 
sign change 
as a function of doping observed in 
experiments in high T$_c$ superconductors. 
In this formulation the Hall constant is given 
by the logarithmic derivative of the Drude weight with respect to the 
particle density.  
This reactive Hall constant approach has recently been extensively 
used in theoretical \cite{giamarchi} and experimental \cite{science}
studies of interacting (and synthetic) quantum systems. 
 
In more generality, both Hall and Edelstein constants, display signatures of Fermi surface topological transitions which are abundant in two-dimensional quantum materials.
Lifshitz transitions with their associated logarithmically-divergent  Van Hove singularities occur in many systems including
cuprates, iron based superconductors, cobaltates, $\text{Sr}_{2} \text{RuO}_{4}$ and
heavy fermions.~\citep{Aoki,Barber,Benhabib,Khan,Coldea,Okamoto,Sherkunov-Chubukov-Betouras,Slizovskiy-Chubukov-Betouras,Yelland}
There is an even more recent surge of interest in higher order Van
Hove singularities \cite{Efremov-Betouras, Chandrasekaran-Shtyk-Betouras-Chamon, LiangFu, Chandrasekaran-Betouras, Chandrasekaran-Betouras-2}. Some of the materials where they have been discovered include $\text{Sr}_{3} \text{Ru}_{2} \text{O}_{7}$
where a higher order ($X_9$ with $n=4$) Van Hove singularity was shown to exist in the presence
of an external magnetic field,~\citep{Efremov-Betouras} while a different types of higher
order Van Hove saddles have been reported in highly overdoped graphene~\citep{Rosenzweig_2020}, the surface of  $\text{Sr}_{2} \text{RuO}_{4}$ \cite{Chandrasekaran-Betouras-3}, kagome metals \cite{Kang, Consiglio, Neupert,Gao} and high-T$_c$ superconductors \cite{Markiewicz}.

In this work, we will  first study the effect of the spin-orbit  interaction on the reactive Hall effect 
of a two-dimensional tight binding model of non-interacting fermions, showing ballistic transport, as a function 
of spin-orbit coupling.  We study its behavior as the Fermi surface evolves as a function of filling. 
We obtain the signatures on  R$_H$ of Fermi surface topological transitions, 
with the associated van Hove singularities in a quantum mechanical description.
Earlier work, using the Boltzmann equation studied these effects 
at a semiclassical level \cite{Kokkinis_etal} and a recent one, on the spin 
and orbital Edelstein effect in a bilayer system with Rashba interaction 
\cite{bilayer}.
Next, we develop a novel formula, in analogy to the reactive Hall constant, 
for the Edelstein effect which describes the appearance of a transverse 
magnetization due to a charge current in a two-dimensional system 
with spin-orbit interaction. 
It would be fascinating to experimentally study the 
reactive Edelstein effect, for instance in synthetic systems \cite{science}, 
as the reactive Hall effect. 

\section{Model}
We consider a generic 
Hamiltonian for fermions on a lattice, where for simplicity we describe the 
kinetic energy term by a one band tight binding model; 
it is straightforward to extend this formulation to a many-band or 
continuum system.
The sites are labeled $l (m)$ along the ${\hat x} ({\hat y})$-direction
with periodic boundary conditions in both directions:

\begin{eqnarray}
H&=&H_0+H_{SO}
\nonumber\\
H_0=\sum_{l,m}&-&t_x e^{i\phi^x(t)}e^{iA_m} c_{l+1,m}^{\dagger}\cdot c_{l,m}
\nonumber\\
&-&t_y e^{i\phi^y_{m+1/2}(t)} c^{\dagger}_{l,m+1}\cdot c_{l,m}+H.c.
\nonumber\\
H_{SO}=\sum_{l,m}
&+&\lambda_x e^{i\phi^x(t)}e^{iA_m} c^{\dagger}_{l+1,m}(-i\sigma^y)c_{l,m}
\nonumber\\
&+&\lambda_y e^{i\phi^y_{m+1/2}(t)} c^{\dagger}_{l,m+1}(i\sigma^x)c_{l,m}
+H.c.
\nonumber\\
l=1,...,L_x&;&~~m=1,...,L_y,~~N=L_x\cdot L_y.
\label{hamiltonian}
\end{eqnarray}
\noindent
where $t_{x,y}$ are the hopping parameters, $\lambda_{x,y}$ the Rashba 
spin-orbit couplings, $\sigma^{\alpha},~\alpha=x,y,z$ 
spin-1/2 Pauli matrices and 
the fermion creation (annihilation) operators are denoted as,
$c^{\dagger}_{l,m}=(c^{\dagger}_{\uparrow l,m}~~ 
c^{\dagger}_{\downarrow l,m})$.

We take a unit lattice constant, electric charge $e=1$ and
$\hbar=1$. We add a magnetic field along 
the ${\hat z}$-direction, modulated by a one component wave-vector-$q$ 
along the ${\hat y}$-direction, generated by the vector potential $A_m$; 
this allows to take the zero magnetic field limit smoothly:
\begin{eqnarray*}
A_m&=&e^{iqm}\frac{iB}{2\sin(q/2)}\simeq e^{iqm}\frac{iB}{q}
\nonumber\\
B_{m+1/2}&=&-(A_{m+1}-A_m)=B e^{iq(m+1/2)}
\label{fieldb}
\end{eqnarray*}

\noindent
(for convenience, we will present the long wavelength limit, 
substituting $2\sin(q/2)\rightarrow q $).
Electric fields along the $\hat x,\hat y$ directions 
are generated by time dependent 
vector potentials:

\begin{eqnarray*}
\phi^{x,y}(t)&=&\frac{E^{x,y}(t)}{iz},~~
\phi^y_{m+1/2}(t)=e^{iq(m+1/2)}\phi^y(t);
\nonumber\\
E^{x,y}(t)&=&E^{x,y} e^{-izt};~~~z=\omega+i\eta~.
\label{phixy}
\end{eqnarray*}

\noindent
Currents are defined through derivatives of the Hamiltonian in $\phi^{x,y}$: 
\begin{eqnarray*}
j^x=-\frac{\partial H}{\partial \phi^x},~~~~~~
j^y_q=-\frac{\partial H}{\partial \phi^y}~,
\label{currents}
\end{eqnarray*}

\noindent
with the paramagnetic parts:
\begin{eqnarray*}
j^x&=&\sum_{l,m} t_xe^{iA_m}ic^{\dagger}_{l+1,m} \cdot c_{l,m}
-\lambda_xe^{iA_m}c^{\dagger}_{l+1,m}\sigma^y c_{l,m} +H.c.
\nonumber\\
j^y_q&=&\sum_{l,m} e^{iq(m+1/2)}(t_yic^{\dagger}_{l,m+1} \cdot c_{l,m}
-\lambda_yc^{\dagger}_{l,m+1}\sigma^y c_{l,m} +H.c.)
\end{eqnarray*}

\section{Reactive Hall response}
We will analyze the reactive Hall response within 
standard linear response theory, 
\begin{eqnarray*}
\langle j^x\rangle&=&\sigma_{j^x j^x}E^x(t)+\sigma_{j^x j^y_q}E^y(t)
\nonumber\\
\langle j^y_q\rangle&=&\sigma_{j^y_q j^x}E^x(t)+\sigma_{j^y_q j^y_q}E^y(t)
\label{respjj}
\end{eqnarray*}
\noindent
closely following the development in \cite{znlp}. 

Reactive - nondissipative - response occurs in a generic uniform 
(without disorder) interacting 
system at zero temperature, in which case the brackets 
$\langle...\rangle$ denote ground state average (\cite{znlp}) 
or at finite temperatures in a uniform noninteracting system 
in which case the brackets $\langle...\rangle$ denote thermal average.
Here we study the reactive Hall constant   
for the Hamiltonian (\ref{hamiltonian}) at finite temperatures 
in the presence of a magnetic field, with the conductivities  
given by,

\begin{eqnarray*}
\sigma_{j^{\alpha}j^{\beta}}&=&\frac{i}{z}
(\langle\frac{\partial^2H}{\partial\phi^{\alpha}\partial\phi^{\beta}}\rangle-
\chi_{j^{\alpha}j^{\beta}}),
\nonumber\\
\chi_{AB}&=&i\int^{\infty}_0 dt e^{izt} \langle [A(t),B] \rangle.
\label{sigma}
\end{eqnarray*}

\noindent
In contrast to the usual derivation of the Hall constant expression, 
we keep the $q-$dependence explicit by converting the 
current-current to current-density correlations using the continuity equation:

\begin{eqnarray*}
\langle j^x\rangle&=&\sigma_{j^x j^x}E^x(t)+\frac{1}{q}\chi_{j^x n_q}E^y(t)
\nonumber\\
\langle j^y_q\rangle&=&-\frac{1}{q}\chi_{n_q j^x}E^x(t)+
(\frac{z}{q})^2\chi_{n_q n_q}\frac{i}{z}E^y(t)~,
\label{respjn}
\end{eqnarray*}

\noindent
with $n_q=\sum_{l,m} (-ie^{iqm}) c^{\dagger}_{l,m}\cdot c_{l,m}$.

At this point we consider the ``screening" (or slow)
response in the ${\hat y}-$direction, by taking the $(q,\omega)$ limits in 
the order $\omega\rightarrow 0$ first and 
$q\rightarrow 0$ last. For $H(\lambda,\mu)$, 
using the following identity,
\begin{equation}
\frac{\partial^2 E_n}
{\partial\mu \partial \lambda}=
\langle n|\frac{\partial^2 H}{\partial\mu \partial\lambda}|n\rangle
-\sum_{m\ne n}\frac{\langle n|\frac{\partial H}{\partial \mu}|m\rangle
\langle m|\frac{\partial H}{\partial \lambda}|n\rangle+H.c.}
{E_m-E_n}
\label{ident}
\end{equation}
\noindent
we arrive at,
\begin{equation}
R_H=-\frac{1}{D}\frac{\partial D}{\partial \rho},
\label{rh}
\end{equation}

\noindent
$\rho$ the electron density.
The Drude weight $D$ can be evaluated by Kohn's expression \cite{kohn} 
and its finite temperatures extension \cite{czp} 
as the second derivative of energy 
eigenvalues $E_n$ with respect to a uniform fictitious flux $\phi_x$,

\begin{equation}
D=\frac{1}{2N}\sum_n p_n
\frac{\partial^2 E_n}{\partial \phi_x^2}
\Bigg|_{\phi_x \rightarrow 0} 
\label{drude}
\end{equation}
\noindent
where $p_n=e^{-\beta E_n}/\sum_n e^{-\beta E_n}$,
$\beta=1/k_B T$ the inverse temperature, $k_B=1$ the Boltzmann constant
and $\rho$ the fermion density.

\bigskip
To evaluate the eigenvalues/eigenstates we diagonalize  
the Hamiltonian (\ref{hamiltonian}) in the presence of the 
fictitious flux $\phi_x$ by transforming to momentum 
${\bf k}=(k_x,k_y)$ space,
\begin{equation*}
\begin{vmatrix}
\epsilon^{(0)}_{\bf k}-\epsilon&\Delta_{\bf k}e^{+i\theta_{\bf k}}\\
\Delta_{\bf k}e^{-i\theta_{\bf k}} &\epsilon^{(0)}_{\bf k}-\epsilon
\end{vmatrix}
=0
\end{equation*}

\begin{equation*}
\epsilon^{(0)}_{\bf k}(\phi_x)=-2t_x\cos( k_x+\phi_x)-2t_y \cos k_y.
\end{equation*}
\begin{equation*}
\Delta_{\bf k}=2\sqrt{\lambda_x^2\sin^2 (k_x+\phi_x)
+\lambda_y^2\sin^2 k_y}
\end{equation*}
\begin{equation*}
\theta_{\bf k}=\tan^{-1} \frac{\lambda_x \sin k_x}{\lambda_y \sin k_y}.
\end{equation*}

\noindent
We obtain 
$H=\sum_{{\bf k}\pm}
\epsilon_{{\bf k}\pm} c^{\dagger}_{{\bf k}\pm}c_{{\bf k}\pm}$
with the two spin states turning by the spin-orbit interaction 
to two chirality eigenstates 
$c_{{\bf k}\pm}=\frac{1}{\sqrt{2}}(
c_{{\bf k}\uparrow}\pm e^{-i\phi_{\bf k}} c_{{\bf k}\downarrow})$
and eigenvalues,
\begin{equation*}
\epsilon_{{\bf k}\pm}(\phi_x)=\epsilon_k(\phi_x)
\pm \Delta_{\bf k}. 
\end{equation*}
\noindent
Finally the reactive Hall constant is given by,
\begin{eqnarray*}
D&=&\sum_{\pm}\int_{-\pi}^{+\pi}\frac{dk_x}{2\pi}
\int_{-\pi}^{+\pi}\frac{dk_y}{2\pi} 
f_{{\bf k}\pm}
\frac{\partial^2 \epsilon_{{\bf k}\pm}}{\partial \phi_x^2} 
\nonumber\\
f_{{\bf k}\pm}&=& \frac{1}{1+e^{\beta(\epsilon_{{\bf k}\pm}-\mu)}}.
\end{eqnarray*}

\begin{figure}[ht]
\includegraphics[angle=0, width=1.0\linewidth]{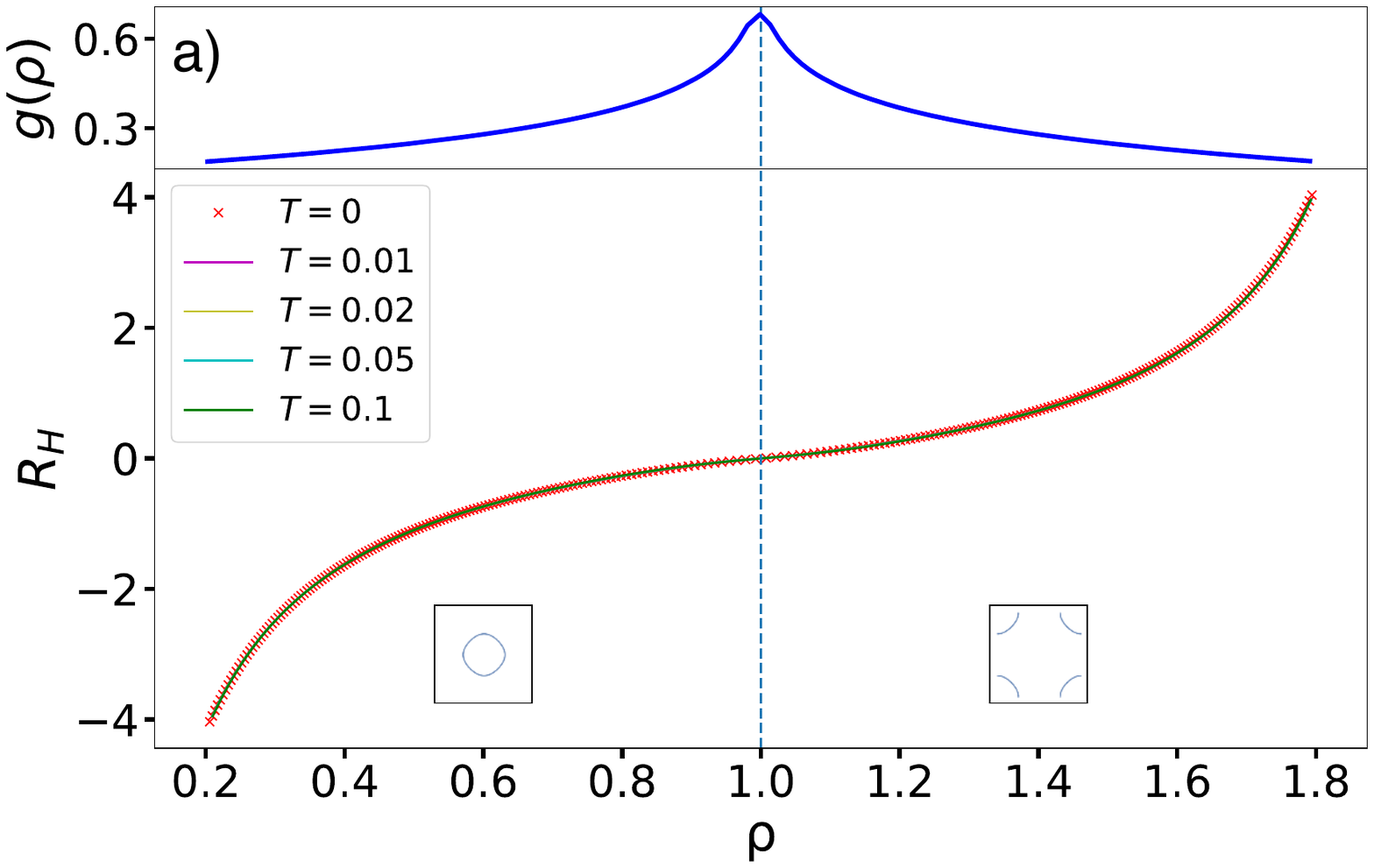}
\includegraphics[angle=0, width=1.0\linewidth]{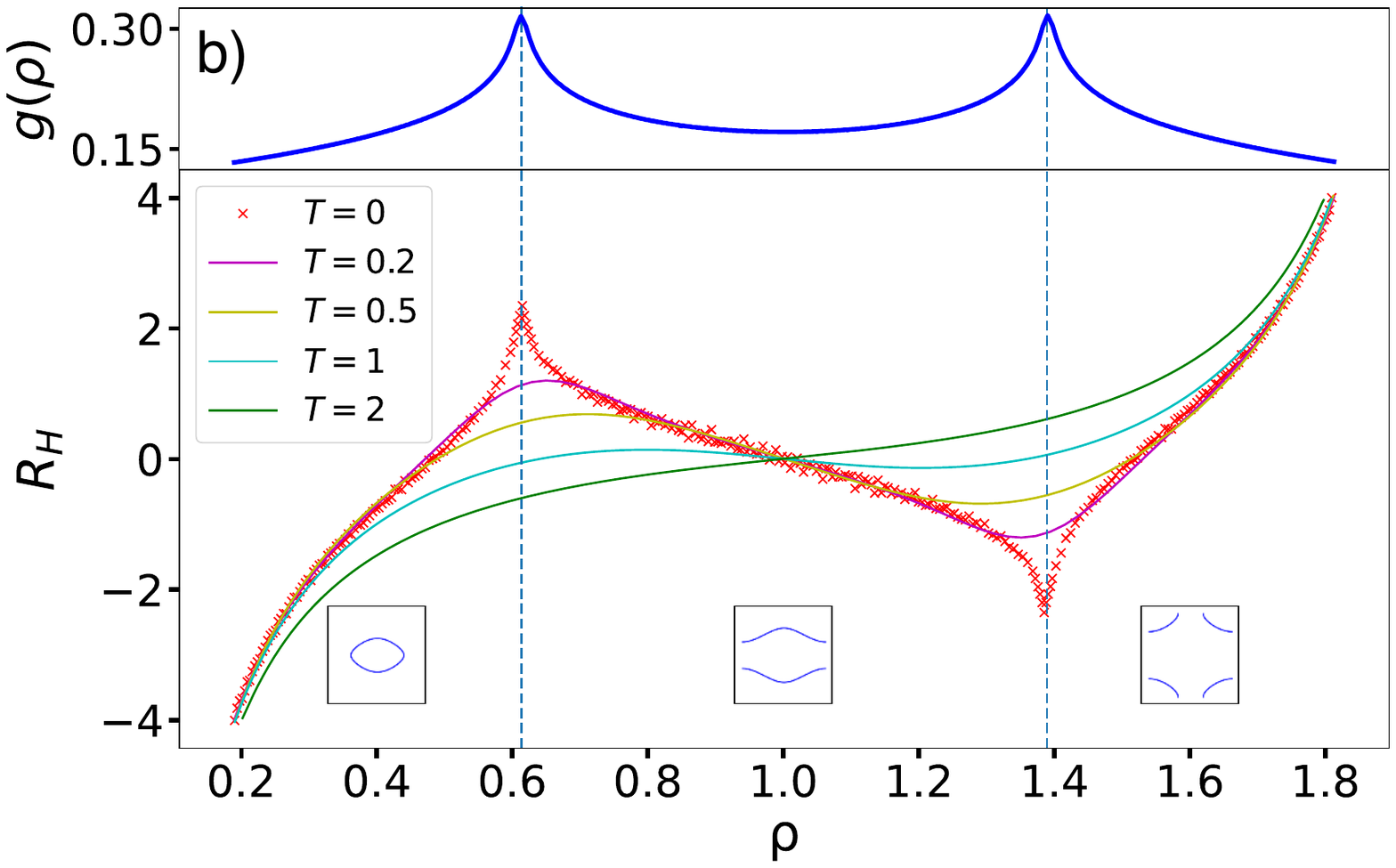}
\caption{The density of states $g(\rho)$ and 
Hall coefficient R$_H$ for cases without SOC, 
(a) t$_x$ =t$_y$=1, (b) t$_x$ =1, t$_y$=2.} 
\label{fig1}
\end{figure}

\begin{figure}[ht]
\includegraphics[angle=0, width=1.0\linewidth]{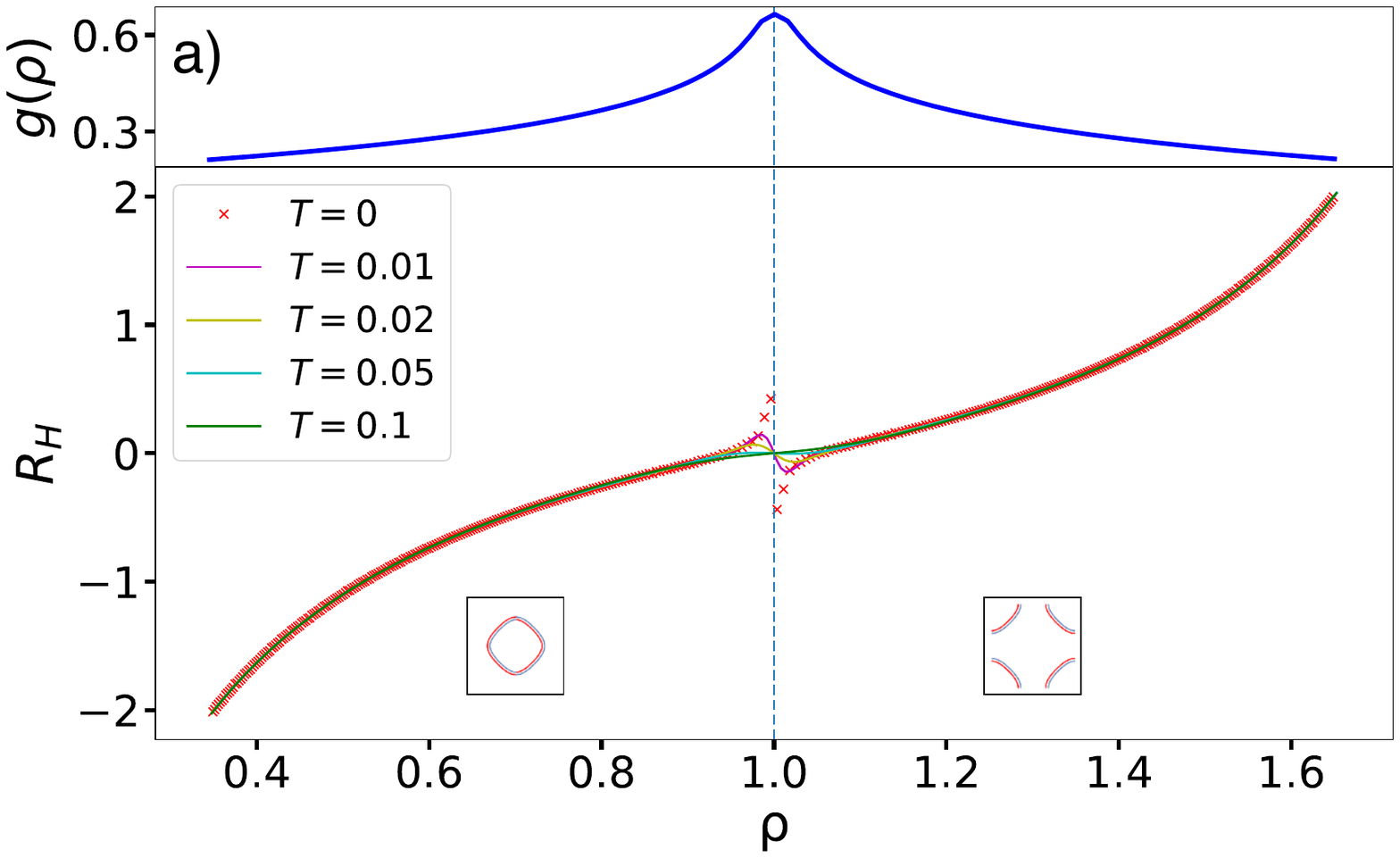}
\includegraphics[angle=0, width=1.0\linewidth]{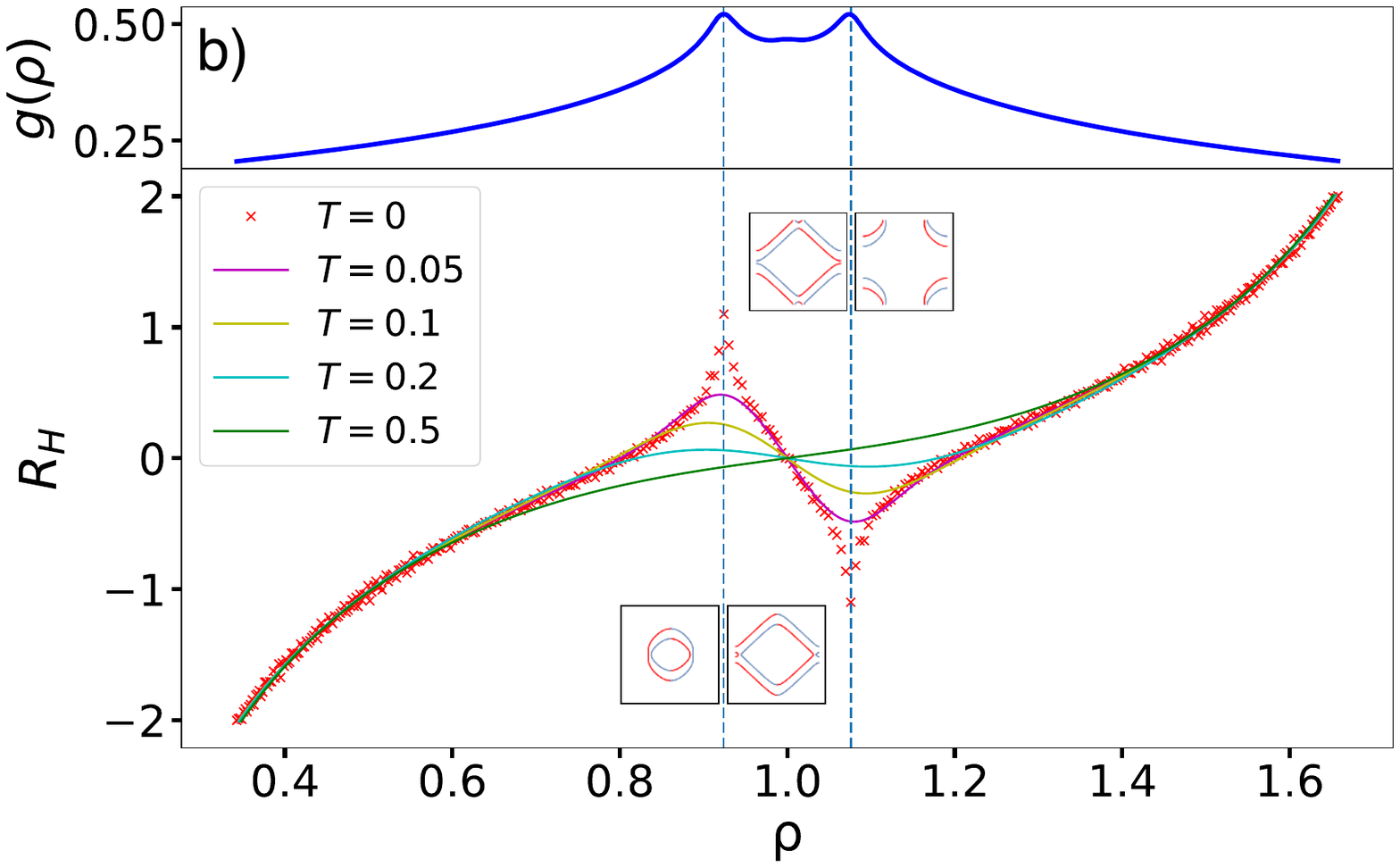}
\includegraphics[angle=0, width=1.0\linewidth]{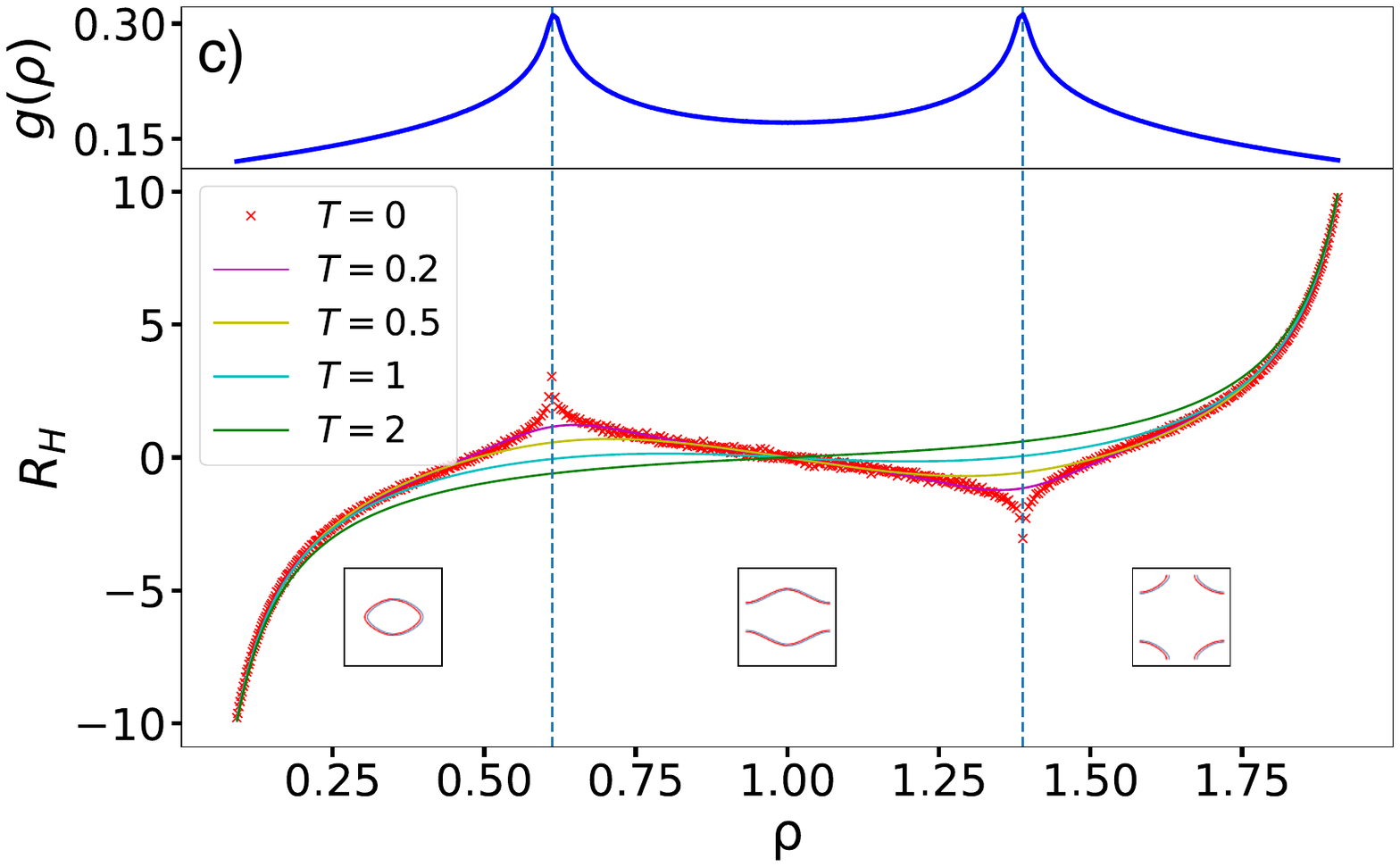}
\caption{The density of states $g(\rho)$ and 
Hall coefficient R$_H$ for cases with SOC, 
(a) t$_x$ = t$_y$= 1 and $\lambda_x = \lambda_y$ = 0.1, 
(b) t$_x$ = 1, t$_y$= 1 and $\lambda_x = 0.1$ $\lambda_y$ = 0.4 and 
(c) t$_x$ = 1,  t$_y$= 2 and $\lambda_x$ = $\lambda_y$ = 0.1.}
\label{fig2}
\end{figure}

Using this equation, we show in Figs.\ref{fig1},\ref{fig2} the 
Hall constant for different
values of the hopping and spin-orbit parameters as a function of 
fermion density $\rho$ in parallel to the density of states,
\begin{equation*} 
g(\rho)=\sum_{\pm}\int_{-\pi}^{+\pi}\frac{dk_x}{2\pi}
\int_{-\pi}^{+\pi}\frac{dk_y}{2\pi} \delta(\mu(\rho)-\epsilon_{{\bf k}\pm}).
\end{equation*} 

In Fig.\ref{fig1}, without any spin-orbit coupling, 
we observe a clear indication in the Hall constant of the van Hove 
singularities. When there is a Fermi surface topological transition, 
there is a discontinuity in the derivative of R$_H$ with respect to $\rho$.
In addition, there is a change of sign when there is a transition 
from an electron-like to a hole-like dispersion. At finite temperature 
the behavior is continuous and the signature of 
the transition from an electron-like to a hole-like dispersion is reflected 
on the monotonic behavior of R$_H$. This goes away at very high temperature 
where R$_H$ has a monotonic behavior independent of the parameters.

\section{Edelstein coefficient}
Along the same line, we can also study the 
Edelstein effect in a two dimensional system with spin-orbit 
coupling where a charge current in the ${\hat x}$-direction induces a 
bulk magnetization density 
$\langle s^y\rangle=\langle 
\frac{1}{N}\sum_{l,m} c^{\dagger}_{l,m} \sigma^y c_{l,m} \rangle $
pointing in the $\hat y$-direction.
The corresponding response is given by,
\begin{equation*}
\alpha_{yx}(\omega)=\frac{1}{i\omega}i \int_0^{\infty} e^{izt}
<[s^y(t),j^x]> dt,~~~z=\omega+i\eta.
\end{equation*}

Here we consider the system (\ref{hamiltonian}) in zero magnetic field 
$A_m=0$ and zero electric field $\phi^y_{m+1/2}(t)=0$ 
along the $\hat y$-direction.
Now we can define a ''reactive Edelstein constant" 
as the prefactor of the $1/\omega$ 
imaginary part of $\alpha_{yx}$,

\begin{eqnarray*}
\frac{D_E}{\omega}&=&
\lim_{\omega\rightarrow 0} 
\Im\alpha_{yx}(\omega)
\nonumber\\
D_E&=&-\sum_n p_n
\sum_m \frac{<n|s^y|m><m|j^x|n>+H.c.}{\epsilon_m-\epsilon_n}
\end{eqnarray*}
\noindent
which using the identity (\ref{ident}) can be written similarly to 
the Kohn formula,
\begin{equation*}
D_E=\frac{1}{N}\sum_n p_n
\frac{\partial^2 E_n}{\partial h^y\partial \phi_x}{\Bigg|_{h^y,\phi_x 
\rightarrow 0}}.
\end{equation*}
\noindent
Here the energy derivatives are over a uniform fictitious flux $\phi_x$
in the ${\hat x}$-direction and a Zeeman field $h^y$ 
along the ${\hat y}$-direction.

For the noninteracting fermion Hamiltonian (\ref{hamiltonian}), 
the imaginary part of the response function becomes,
%\onecolumngrid
\begin{eqnarray*}
D_E&=&
\lim_{\omega\rightarrow 0} 
\sum_k (2\lambda_x\cos k_x \cos^2 \phi_{\bf k} )
(f_{{\bf k}+}-f_{{\bf k}-})\cdot
\nonumber\\
&&[
\frac{\omega-\delta_{\bf k}}{(\omega-\delta_{\bf k})^2+\eta^2}-
\frac{\omega+\delta_{\bf k}}{(\omega+\delta_{\bf k})^2+\eta^2}]
\nonumber\\
\delta_{\bf k}&=&\epsilon_{{\bf k}+}-\epsilon_{{\bf k}-}.
\end{eqnarray*}
%\twocolumngrid

As discussed above, this expression for the imaginary part of the 
response function, can be recast in the form,
\begin{equation}
D_E=\sum_{\pm}\int_{-\pi}^{+\pi}\frac{dk_x}{2\pi}
\int_{-\pi}^{+\pi}\frac{dk_y}{2\pi} f_{{\bf k}\pm}
\frac{\partial^2 \epsilon_{{\bf k}\pm}}
{\partial h^y \partial \phi_x}\Bigg|_{h^y,\phi_x \rightarrow 0}.
\label{de}
\end{equation}
\noindent
To evaluate the derivative $\frac{\partial^2 \epsilon_{{\bf k}\pm}}
{\partial h^y \partial \phi_x}|_{h^y,\phi_x \rightarrow 0}$
we add in the Hamiltonian (\ref{hamiltonian}) a magnetic field 
in the $\hat y$-direction ($+h^y s^y$), modifying the eigenvalues as,
\begin{eqnarray*}
\epsilon_{{\bf k}\pm}&=&\epsilon^{(0)}_{\bf k}\pm 2
\sqrt{(\lambda_x\sin (k_x+\phi_x)-h^y/2)^2
+\lambda_y^2\sin^2 k_y}
\noindent\\
&=&\epsilon^{(0)}_{\bf k}(\phi_x)\pm \Delta_{\bf k} (\phi_x,h^y),
\end{eqnarray*}
\noindent 
and finally obtaining,
\begin{equation}
\frac{\partial^2 \epsilon_{{\bf k}\pm}}{\partial h^y \partial \phi_x}=
(\mp)\frac{(2\lambda_x\cos k_x)(2\lambda_y \sin k_y)^2}
{\Delta_{\bf k}^3|_{h^y=0,\phi_x=0}}.
\label{de2}
\end{equation}

Note from (\ref{de}) and (\ref{de2}) that inverting $\lambda_x \rightarrow -\lambda_x$,
$D_E \rightarrow -D_E$, while $D_E$ remains invariant changing the sign of 
$\lambda_y$.

As we are interested in probing Fermi surface topological transitions, the derivatives of $D_E$ with respect to chemical potential, trace FS topological transitions.
At the points where the curvature of $D_E$ as a function of fermion density $\rho$ changes sign, we've got the relation:
\begin{equation}
\frac{d^2D_E/d\mu^2}{dD_E/d\mu} = - \frac{d^2\mu/d\rho^2}{(d\mu/d\rho)^2}. 
\end{equation}

The ratio of the derivatives of $D_E$ on the left-hand side, probe Fermi surface topological transitions because at T=0 the integrations are along Fermi surface paths within the first BZ:
\begin{widetext}

\begin{eqnarray}
\frac{d^2D_E}{d\mu^2}=(\mp) \sum_{\pm} \frac{\partial}{\partial\mu} \int_{-\pi}^{+\pi}\frac{dk_x}{2\pi}
\int_{-\pi}^{+\pi}\frac{dk_y}{2\pi} 
\frac{(2\lambda_x\cos k_x)(2\lambda_y \sin k_y)^2}
{\Delta_{\bf k}^3|_{h^y=0,\phi_x=0}} \delta(\mu- \epsilon_{{\bf k}\pm})  \\
\frac{dD_E}{d\mu}=(\mp) \sum_{\pm} \int_{-\pi}^{+\pi}\frac{dk_x}{2\pi}
\int_{-\pi}^{+\pi}\frac{dk_y}{2\pi}
\frac{(2\lambda_x\cos k_x)(2\lambda_y \sin k_y)^2}
{\Delta_{\bf k}^3|_{h^y=0,\phi_x=0}} \delta(\mu- \epsilon_{{\bf k}\pm}) 
\end{eqnarray}

\begin{figure}[h]
\centering
\includegraphics[width=0.45\textwidth]{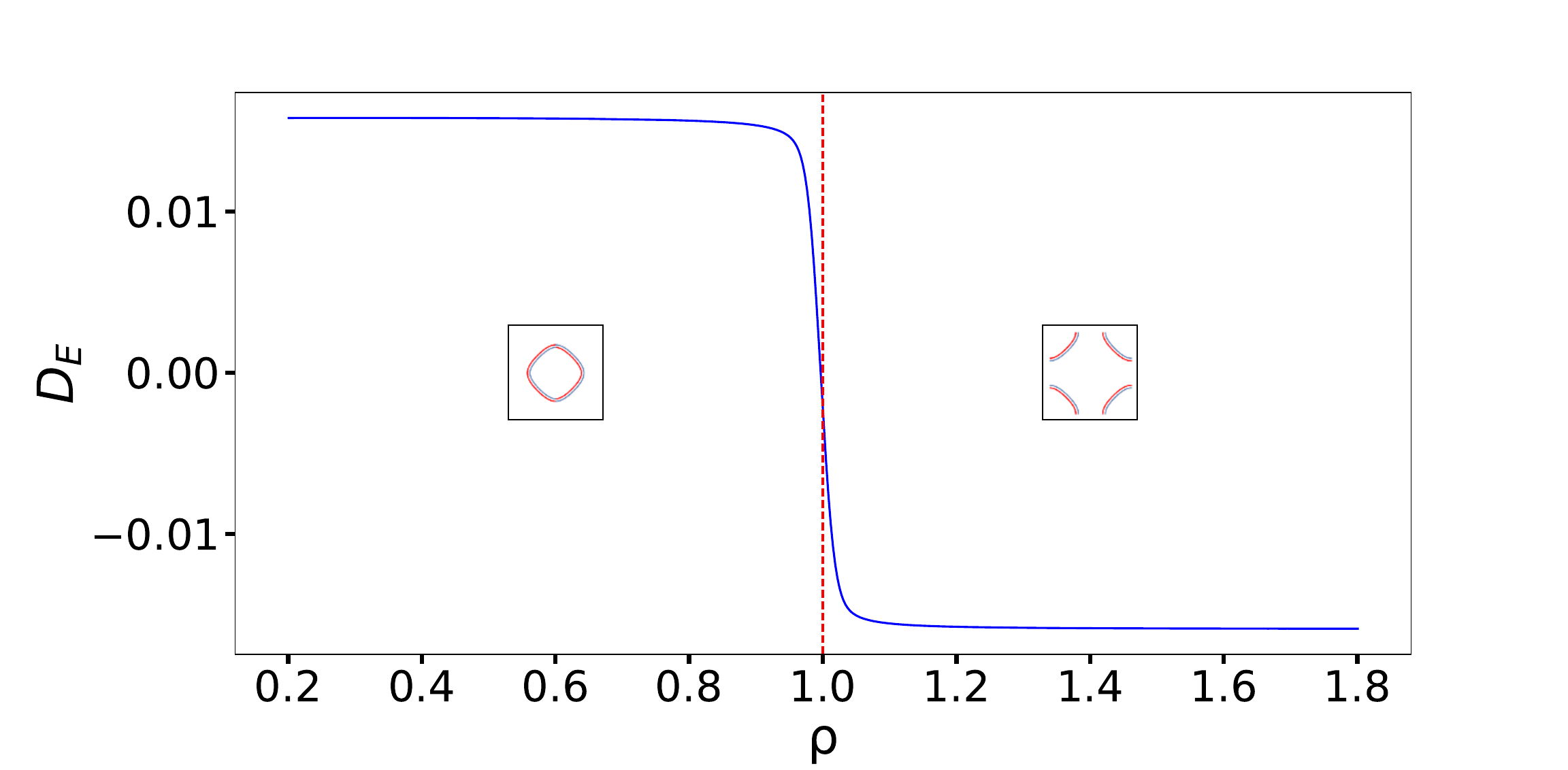}
\includegraphics[width=0.45\textwidth]{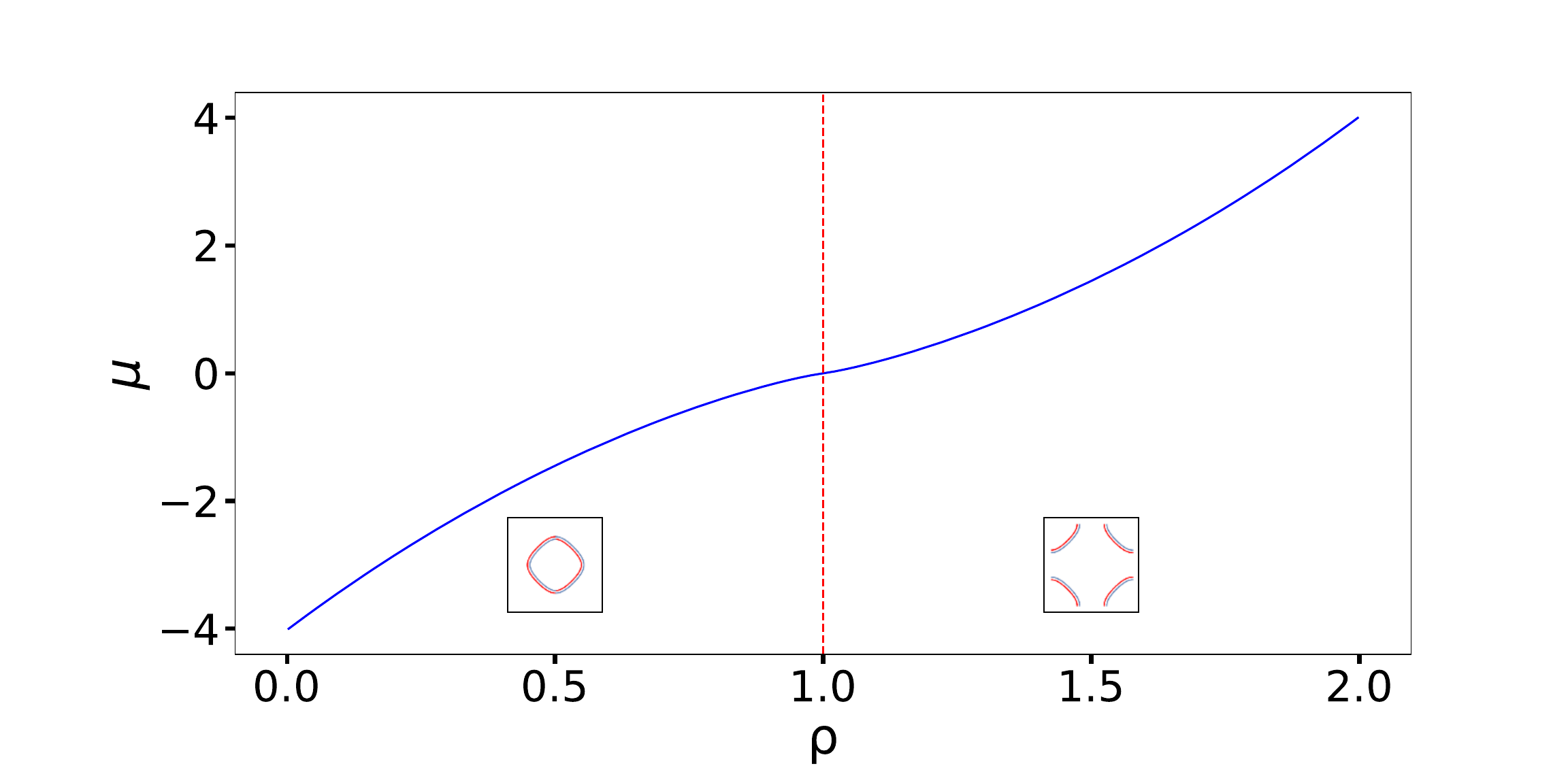}
\caption{Left-hand side: the Edelstein constant $D_E$ as a function 
of fermion density, for t$_x$ = t$_y$= 1 and $\lambda_x = \lambda_y$ = 0.1, 
right-hand side: the chemical potential as a function of fermion density 
for the same parameters.}\label{fig3}
\includegraphics[width=0.45\textwidth]{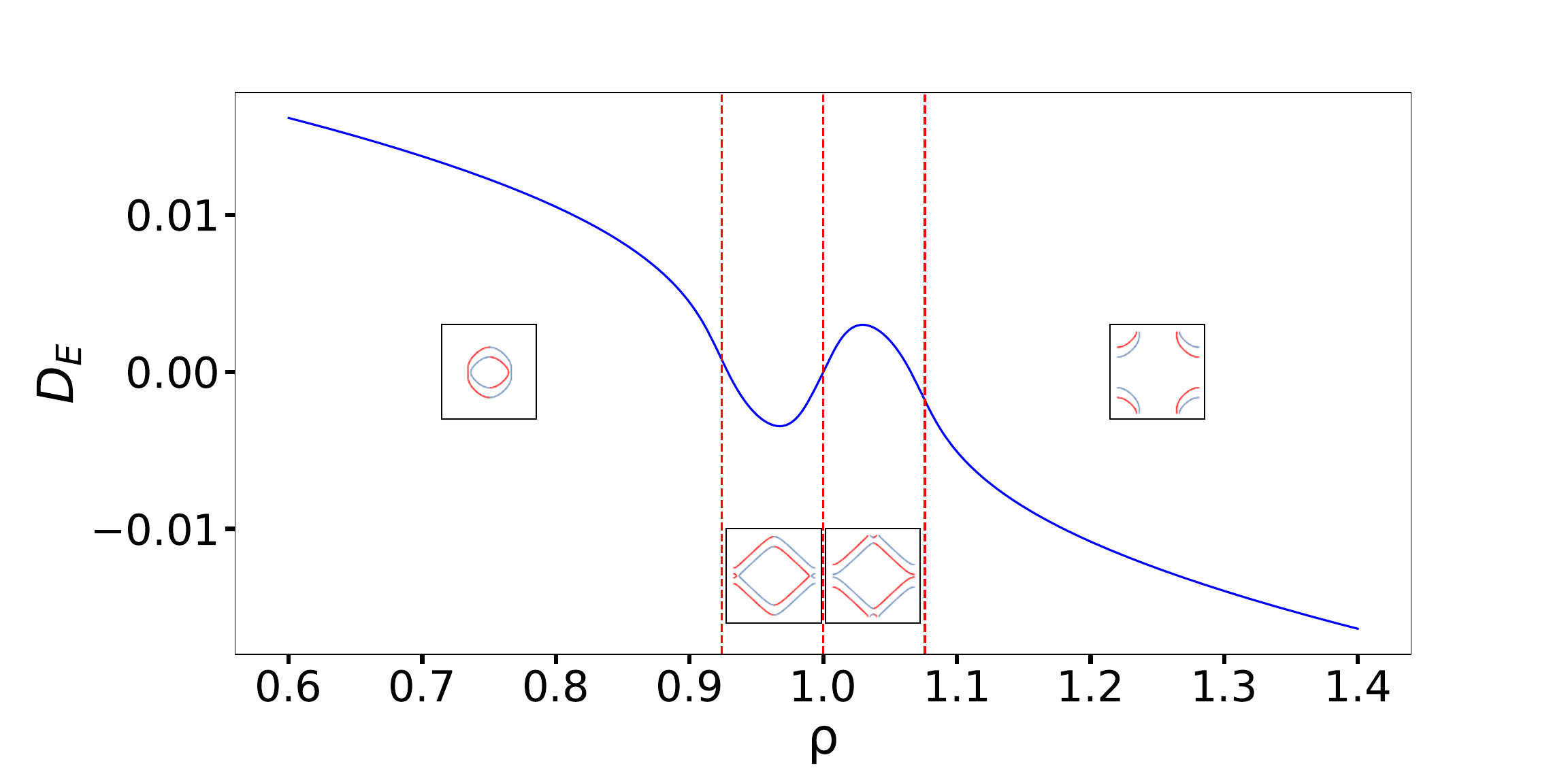}
\includegraphics[width=0.45\textwidth]{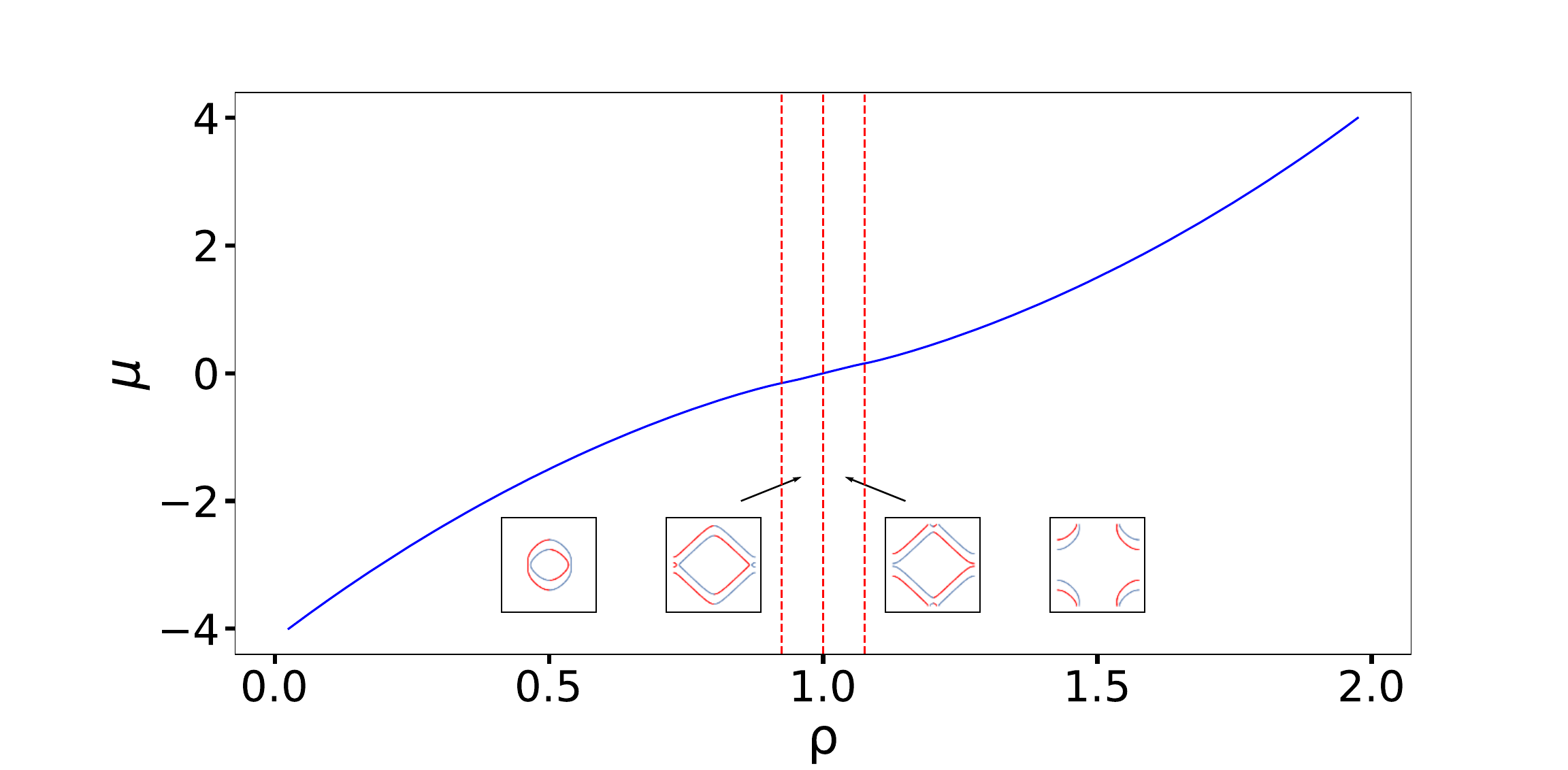}
\caption{Left-hand side: the Edelstein constant $D_E$ as a function 
of fermion density for t$_x$ = 1, t$_y$= 1 and $\lambda_x = 0.1$ 
$\lambda_y$ = 0.4, right-hand side: the chemical potential as a function 
of fermion density.}\label{fig4}
\includegraphics[width=0.45\textwidth]{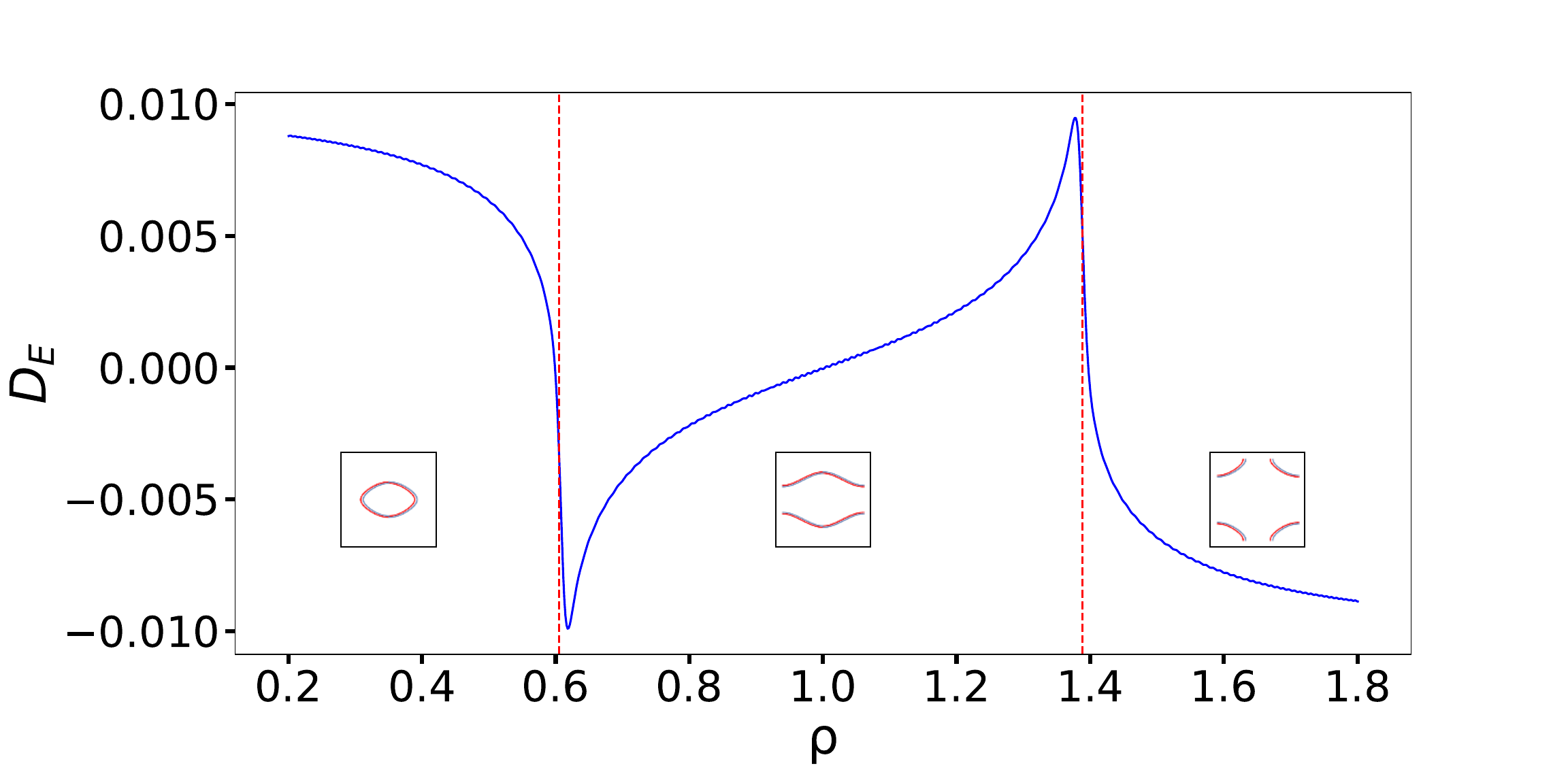}
\includegraphics[width=0.45\textwidth]{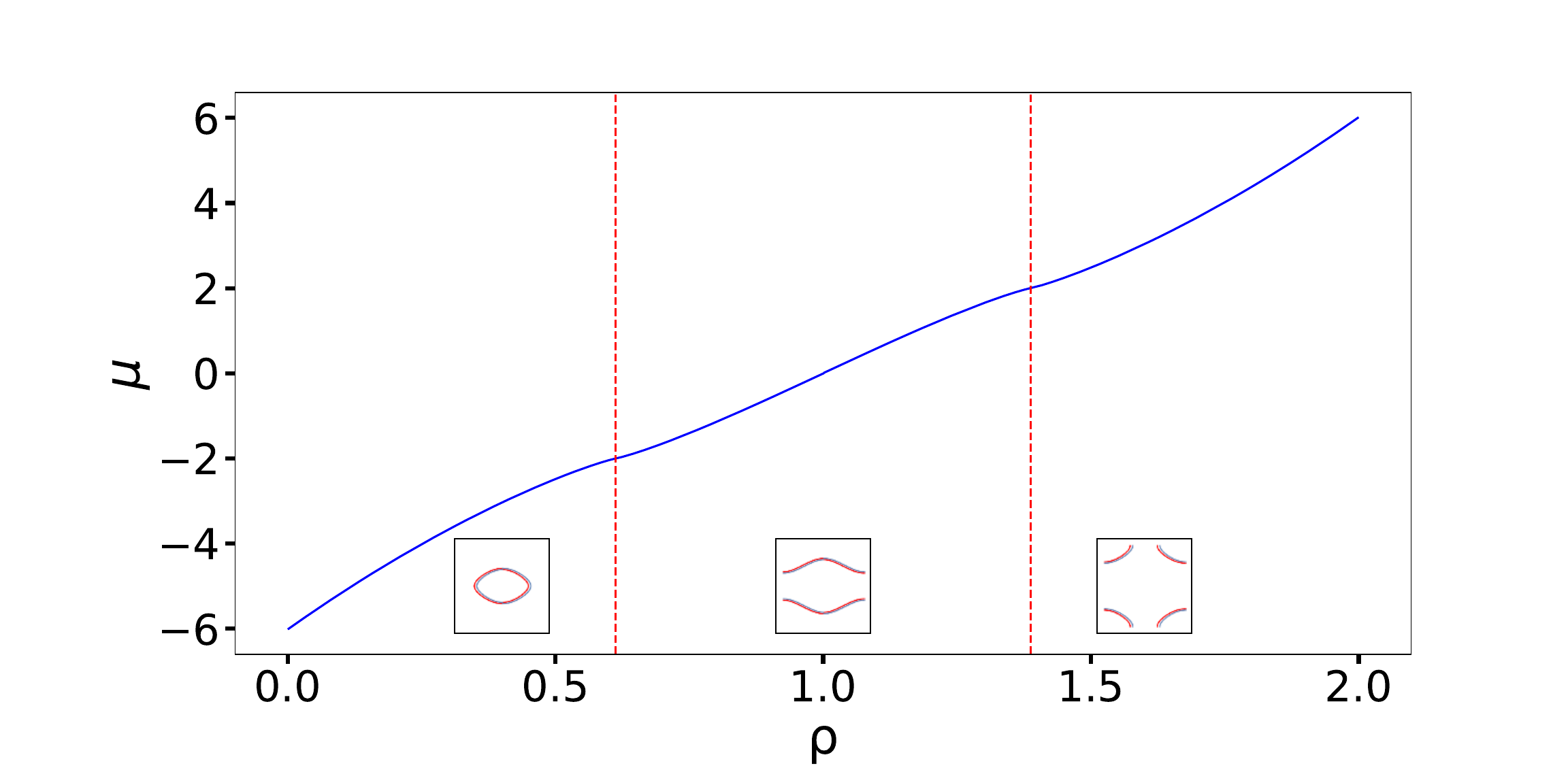}
\caption{Left-hand side: the Edelstein constant $D_E$ as a function 
of fermion density now for t$_x$ = 1,  t$_y$= 2 and $\lambda_x$ = 
$\lambda_y$ = 0.1 with the corresponding chemical potential as a 
function of density.}\label{fig5}
\end{figure}
\end{widetext}

\begin{figure}[ht!]
\includegraphics[angle=0, width=1.0\linewidth]{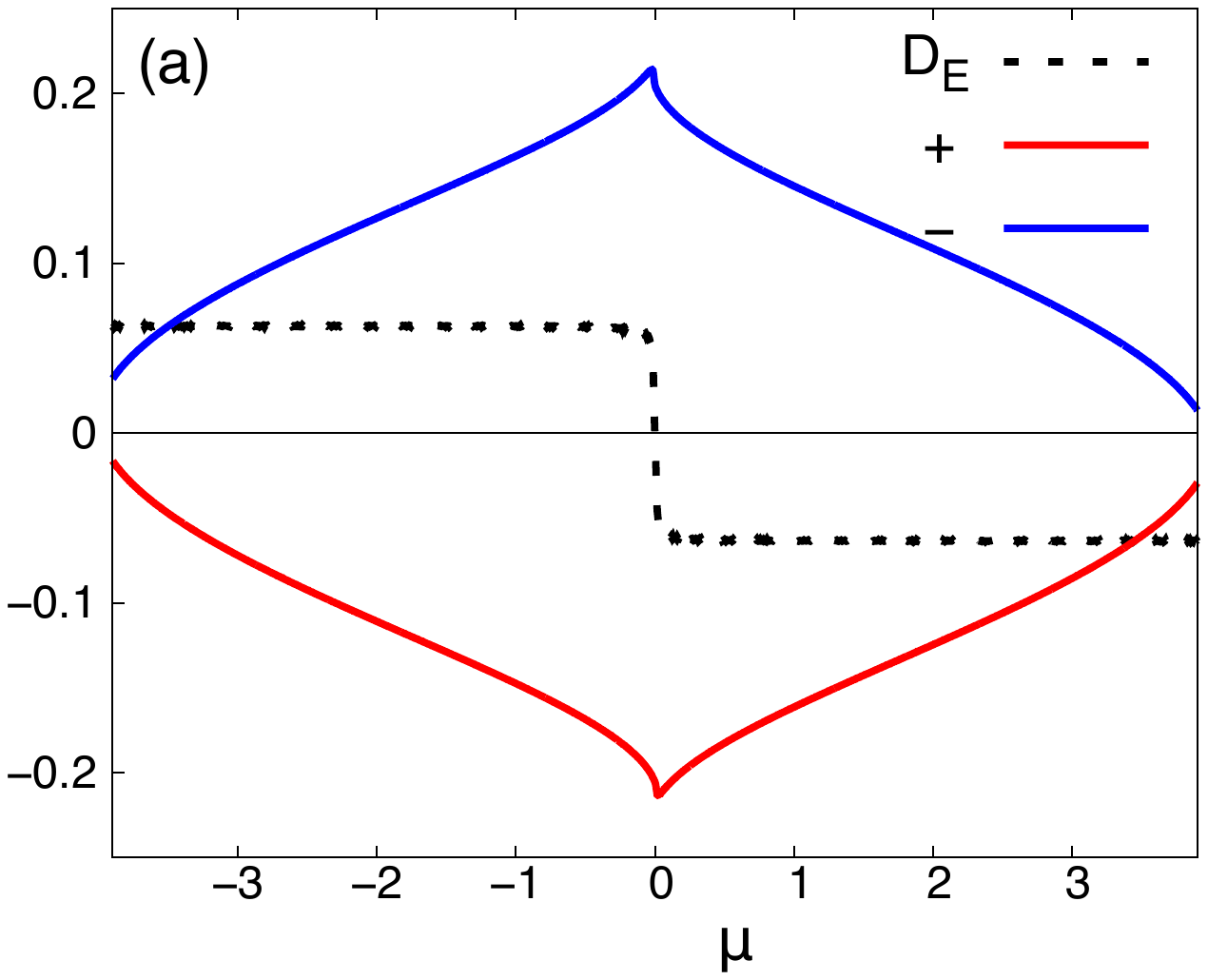}
\includegraphics[angle=0, width=1.0\linewidth]{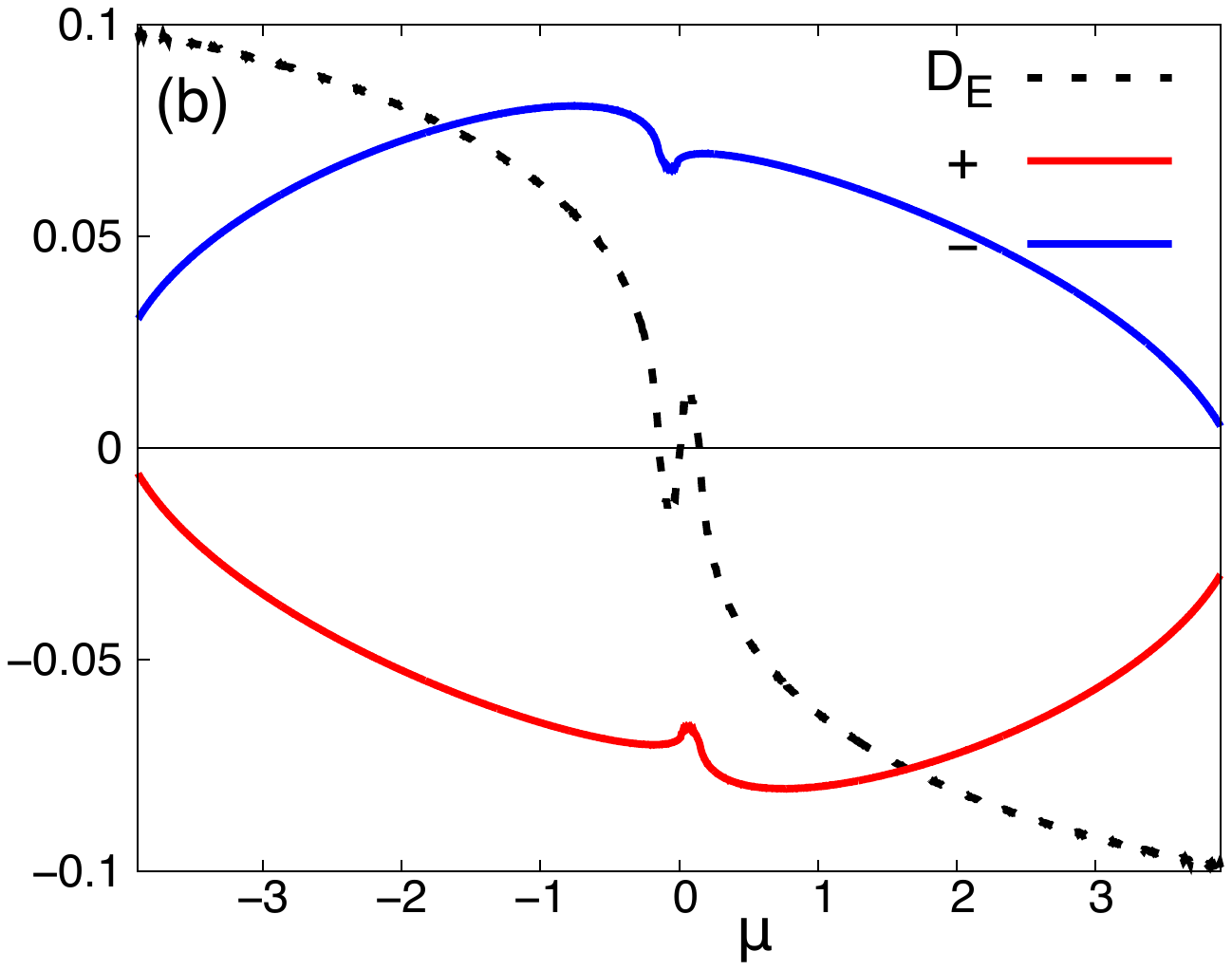}
\includegraphics[angle=0, width=1.0\linewidth]{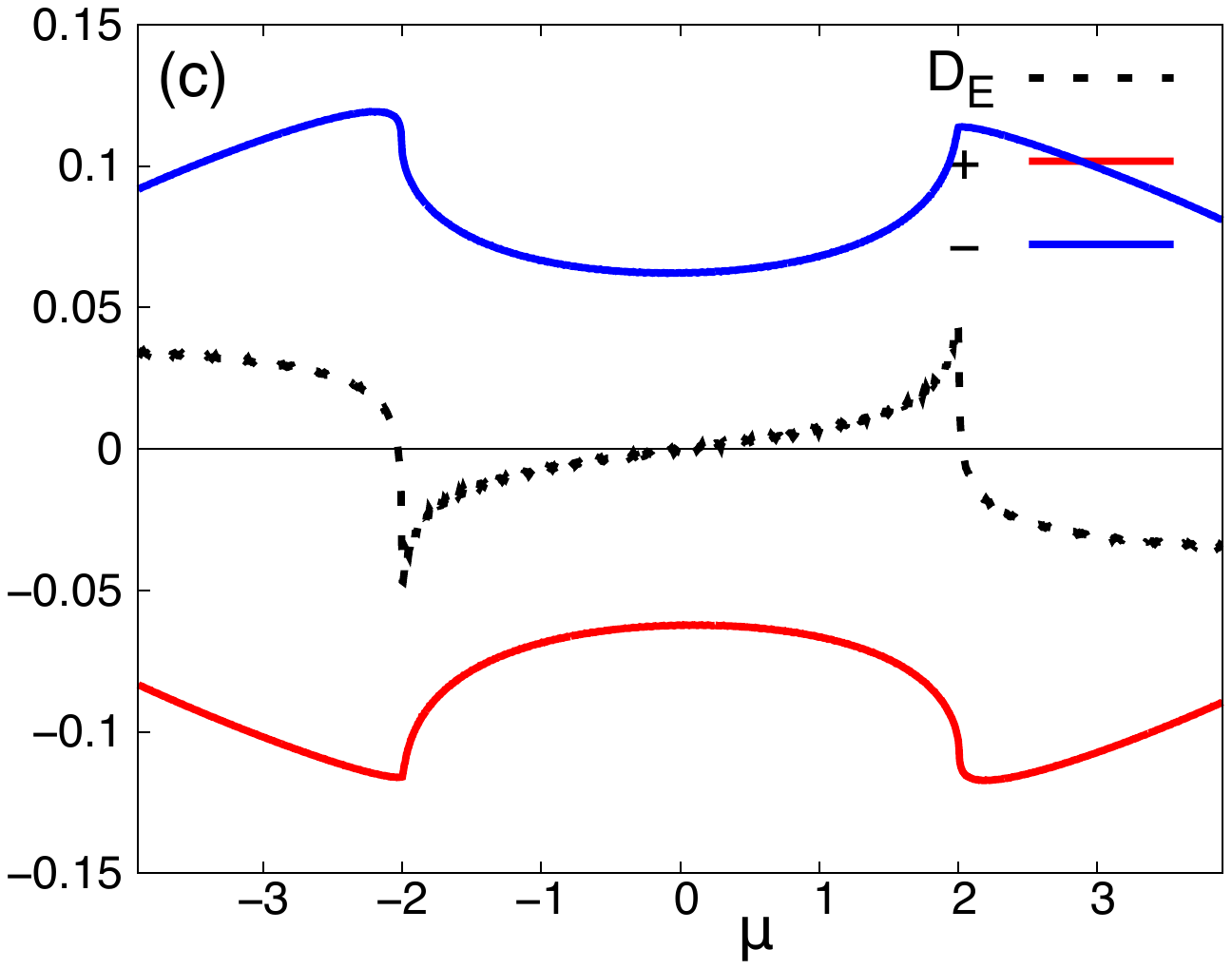}
\caption{The two chirality ($\pm$) contributions to the Edelstein constant
(\ref{de})
and the total $D_E$ (as in Figs.\ref{fig3},\ref{fig4},\ref{fig5}, 
multiplied by a factor of 4 for clarity),
(a) t$_x$ = t$_y$= 1 and $\lambda_x = \lambda_y$ = 0.1,
(b) t$_x$ = 1, t$_y$= 1 and $\lambda_x = 0.1$ $\lambda_y$ = 0.4 and
(c) t$_x$ = 1,  t$_y$= 2 and $\lambda_x$ = $\lambda_y$ = 0.1.}
\label{fig6}
\end{figure}

Fig.\ref{fig3} shows $D_E$ at
zero temperature as a function of fermion 
density for $t_x=t_y=1$ and $\lambda_x = \lambda_y = 0.1$. We see that $D_E$ changes sign as we go from a closed to an open Fermi surface.
As we noted before, this behavior captures the transition from electron-like to hole-like pocket.
Similarly, Fig.\ref{fig4} illustrates the same physics for the case 
of anisotropic spin-orbit coupling, while in Fig.\ref{fig5} the chosen 
parameters correspond to anisotropic strength of hopping elements.
In Fig.\ref{fig6} we show separately the opposite contributions 
of the two chiralities 
to $D_E$ in (\ref{de}). We find that $D_E$ is the result of a large 
cancellation between the two components each of which however keeps track of 
the FS topological transitions.

The low density limit ${\bf k} \rightarrow 0$ can be treated analytically 
for the isotropic parameters $t_x=t_y=t,~~\lambda_x=\lambda_y=\lambda$.
With $k_x=k \cos \phi,~~k_y=k\sin \phi$ we obtain,
$\epsilon_{k\pm}=-4t+tk^2 \pm 2\lambda k$,

\begin{eqnarray}
D_E&=&\sum_{\pm} 
\frac{1}{(2\pi)^2}\int_0^{\infty} dk k f_{k\pm} \int_0^{2\pi} d\phi\cdot 
\nonumber\\
&(\mp)&\frac{8\lambda^3 (1-k^2\cdot \cos^2 \phi/2)k^2\sin^2 \phi}
{8\lambda^3 k^3}
\nonumber\\
&=&-\frac{1}{4\pi}\int_0^{\infty} dk (f_{k+}-f_{k-})(1-\frac{k^2}{8})
\end{eqnarray}

\noindent 
At $T=0$ the upper cutoff of the integrals becomes 
$k_{\pm c}=\mp \lambda +\sqrt{(4t+\mu)^2+\lambda^2}$, 
\begin{equation}
D_E=-\frac{1}{4\pi}[
(k_{+c}-\frac{k^3_{+c}}{24})
-(k_{-c}-\frac{k^3_{-c}}{24})].
\end{equation}

\section{Conclusions}
In this work we study the effect of spin-orbit interaction 
on the reactive Hall constant $R_H$ within a noninteracting 
fermion Hamiltonian model. 
In particular we show the role of van Hove singularities in 
the density of states as cusps in the density dependence of $R_H$.
Furthermore, we introduce the concept of reactive Edelstein constant,
an effect in spin-orbit coupled systems and derive an expression analogous to 
the Kohn formula for the reactive response.

The main open question is the effect of interactions, introduced 
e.g. by a Hubbard $U$ term in the Hamiltonian. 
An emerging Mott gap, typically at half filling, leads to a vanishing $D$
as Kohn pointed out, the criterion for a 
Mott metal-insulator transition.
Thus the interaction effects in combination with the presence of (higher-order) 
van Hove singularities can have drastic consequence on 
phase formation \cite{Zervou_Efremov_Betouras_2023}.
At first, this can be studied by a mean field approach \cite{mf1,mf2} 
which essentially leads to effective single particle states allowing 
a direct computation of $R_H$ and $D_E$. 
Next, the interaction effects can be studied by 
either numerical simulations or many-body theory techniques 
\cite{Zervou_Efremov_Betouras_2023}. 
An interesting question is the temperature dependence of the Edelstein 
constant. While it is known that the Drude weight vanishes at any 
finite temperature in an interacting system evidence of dissipation due 
to interactions, the temperature dependence of $D_E$ is not clear.
\\

\subsection{Acknowledgements}
We are grateful to the UK Engineering and Physical Sciences Research Council 
(ESPRC)
for funding via Grant No EP/T034351/1 (JJB). 
X.Z. thanks the Condensed Matter group of the Physics Department at the 
University of Loughborough for the hospitality and fruitful discussions during his visit.

\end{document}